\documentstyle[preprint,aps]{revtex}
\tightenlines

\begin{document}
\title{
Construction of transferable spherically-averaged electron potentials}

 \author{K. Stokbro, N. Chetty$^1$, K. W. Jacobsen, and J. K. N\o
rskov}
\address{
Center for Atomic-scale Materials Physics and Physics Department, \\
Technical University of Denmark, DK 2800 Lyngby, Denmark \\
$^1$ Department of Physics, Brookhaven National Laboratory \\
 Upton, NY\ \ 11973, USA}
\narrowtext
\maketitle

\begin{abstract}

A new scheme for constructing approximate effective electron potentials
within density-functional theory is proposed. The scheme
consists of calculating the effective potential for a series of
reference systems, and then using these potentials to construct the
potential of a general system. To make contact to the reference system
the neutral-sphere radius of each atom is used. The scheme can simplify
calculations with partial wave methods in the atomic-sphere or
muffin-tin approximation,
 since potential parameters can be precalculated and
then for a general system obtained through simple interpolation
formulas.  We have applied the scheme to construct electron potentials
of phonons, surfaces, and different crystal structures of
silicon and aluminum atoms, and found excellent agreement with the
self-consistent effective potential. By using an approximate
total electron density obtained from a superposition of atom-based densities,
the energy zero of the corresponding effective potential can be found
and the energy shifts in the mean potential between inequivalent
atoms can therefore be directly estimated.
This approach is shown to work well for surfaces and phonons of silicon.
\end{abstract}

One route that seems promising in order to construct computationally
efficient 'ab initio' schemes for calculating total energies and forces
of solids is to exploit the variational properties of
density-functional theory\cite{hohenberg}. We have  shown earlier how
the total electron density can be decomposed into a superposition of
transferable atom-based densities for metals and
semiconductors\cite{chetty,si}.  When such densities are used to generate
an input density for the Harris functional\cite{harris,haydock},
excellent total energies are obtained for surfaces, phonons and
structural differences\cite{al,si} due to the fact that the Harris
functional is stationary in the density.  In this way the
self-consistency loop is avoided.
It is the purpose of the present report to show how the the variational
nature of density-functional theory can be exploited even further by
working with both approximate densities and
potentials simultaneously.

The Hohenberg-Kohn density functional can be generalized to a
functional $E[n,v]$ which depends on both the density $n$ and the
potential $v$\cite{haydock,jacobsen} and which is stationary with
respect to independent variations of the density and the potential. The
general functional can be written
\begin{equation}
E[n,v]= \sum_{\alpha} \epsilon_\alpha[v] - \int n({\bf r}) v({\bf r})  d{\bf r}
+ E_{el}[n] + E_{xc}[n],
\end{equation}
where $\epsilon_\alpha$ denotes the eigenvalues generated by the potential
$v$, and where $E_{el}[n]$ and $E_{xc}[n]$ is the electrostatic and
exchange-correlation energy functionals, respectively. If the
potential is restricted
to be a functional of the density, the Hohenberg-Kohn functional or the
Harris functional appear as special cases\cite{jacobsen}. The
stationary property of the general functional with respect to
variations in the potential can be utilized to construct efficient
schemes for evaluation of total energies and Hellman-Feynman forces\cite{si}.
In the following we shall describe one such scheme which has its root
in the effective-medium theory\cite{jacobsen}. The scheme applies to
situations in which the kinetic energy can be calculated within the
muffin-tin or atomic-sphere approximation (ASA)  with spherically-symmetric
potentials within the atomic spheres. The idea is to use
self-consistently calculated potentials from a series of reference
systems which we choose here to  be a bulk crystal with varying lattice
constant. For a given atom in a general system the potential within
the atomic sphere  around the atom is then approximated by the
potential in the reference system with an appropriate lattice
constant. The lattice constant of the associated reference system is
determined by the
requirement that a neutral sphere around the atom should have the same
radius in the system under study and  in the reference system.

The neutral-sphere  is
a sphere  containing 3 (4) electrons in the case of aluminum
(silicon) in the pseudo-potential scheme.
If the approach described here is combined with the density
construction of Ref.~\cite{chetty} where the total electron density, $n(r)$,
is approximated by a superposition of atom-based densities, $\Delta n_{atom}$,
 positioned at each atomic site, ${\bf R}_i$,
\begin{equation}
n( {\bf r}) = \sum_i \Delta n_{atom}(| {\bf r}- {\bf R}_i|),
\end{equation}
then the NS radius can be obtained directly, or
simple interpolation formulas can be made from which the NS
radius can be obtained with high accuracy\cite{al,si}.

We have used this scheme to calculate the effective potential
of silicon and aluminum atoms in different configurations.
For silicon we use the diamond structure and for aluminum
the fcc structure as reference systems. The lattice parameter is regarded as a
parameter
which can be varied in order to find a good approximation for the potential.
To calculate the effective potentials we use a
self-consistent plane-wave pseudo-potential program, with a
12 Rydberg cutoff for the plane-wave basis set. With this cutoff, we get the
lowest energy configuration of silicon to be the diamond lattice with
lattice constant 10.17 $a_{0}$, and for aluminum the fcc lattice with
lattice constant 7.48 $a_{0}$.

As test systems for silicon we consider  the diamond longitudinal phonon at the
X point
(denoted LAO(X) and frozen at the displacement 0.02 in units of the lattice
constant), the diamond (100) surface, and the fcc structure with a
lattice constant of 7.18 $a_{0}$.
And similarly for aluminum, we use the fcc longitudinal phonon at the X point
(denoted L(X) and displacement 0.02), the fcc (100) surface, and the diamond
structure with a
lattice constant of 11.05 $a_{0}$.  These six
structures cover the two elements in quite different surroundings, and if the
potential construction works for these situations a high degree of
transferability
is guaranteed.

As the first test system, we consider silicon in the fcc structure.
As noted in the introduction the way we make
contact between the test system and the reference
system is through the neutral-sphere (NS) radius.
 In the fcc structure the NS radius
is almost equal to the Wigner-Seitz (WS) radius, while the NS radius is
substantially
smaller than the WS radius in the diamond structure. This difference is due
to the large regions in the diamond lattice which contain almost no charge
and therefore are not included in the neutral sphere. In Fig.~1 we show the
local part of the self-consistent effective potential of the fcc test system
compared to that of the
reference system with the same NS and WS radius as the fcc test system,
respectively. Clearly
the potential of the reference system chosen according to the NS criterion
gives by far the best approximation to the fcc potential.

In order to quantify the difference in the potentials, we introduce
the r.m.s. error $\sigma$, defined by
 \begin{equation}
 \sigma^2  = \frac{3}{s_w^3} \int_0^{s_w} [\bar{v}(r)-\bar{v}^{ref}(r)]^2 r^2
dr,
 \end{equation}
where $s_w$ is the WS radius of the test system, and the
two potentials are aligned such that they have the same average
within the sphere.
To get an estimate of the error in the potentials
due to the finite plane-wave basis set, we have compared one of the reference
potentials of silicon with that of an 18 Rydberg calculation. We
find a r.m.s. error of $\sigma = 0.06$~eV, so this is the level of accuracy
we ideally can obtain.

In Fig.~2 we show the r.m.s error between the potential of the fcc
test system and the diamond reference system as a function of the WS radius
(i.e. as a function of lattice constant) of the diamond lattice.
We see that optimally the reference system should be chosen with a WS
radius of 3.18 $a_{0}$. We note that this is not at all
close to the WS radius in the fcc test system ($s_w = 2.806$ $a_{0}$).
In the ASA the reference diamond structure is often embedded in a bcc structure
with twice the number of spheres where only half of them
contain an atom, and the others are empty.
For this construction the sphere radius is a factor $2^{1/3}$ smaller
than the WS radius, and the reference system where this smaller sphere
equals the fcc WS radius is given by $s_w = 3.53$ $a_{0}$, which still
is far from the the optimal reference system.
However the reference diamond lattice with the same
NS radius as the fcc test system, has a WS radius of 3.178 $a_{0}$,
and is therefore almost exact the optimal reference system. This is
not just a coincidence, since for all test systems investigated we have
found this to be the case.

In Table~1 we show the minimal r.m.s. error between the potential of the
six test systems and the reference system, compared to the error when
the reference system
is chosen to have the same NS radius or the same WS radius as the test
system.
It is evident from the table that using the reference potential chosen
according to the NS  criterion is almost optimal, while the WS criterion
is far from optimal. It should be noted that if the integral in Eq.~(2)
were  done within the neutral sphere instead of within the WS sphere,
the NS criterion would give even smaller errors.

In order to estimate how much the errors induced by the  approximations for the
potential will affect the total energy, we show in Table~2
the value of LMTO potential parameters\cite{lmto} for the three potentials of
Fig.~1. The potential parameters are calculated by solving the radial
Schr\"{o}dinger equation at a fixed energy for the s and p angular component
($\epsilon_{\nu}=\{13.22,18.96\}$), with the energy chosen  at  the
center of gravity of the
occupied part of the fcc band. The accuracy of the
potential parameters directly reflects the accuracy of the corresponding
one-electron bands, and thereby the one-electron band energy.
As seen from Table~2 the  potential parameters obtained with the NS
criterion are in excellent agreement with the self-consistent
parameters, while the potential parameters obtained with the WS criterion
are more than 10 percent off.

Until now we have aligned the average potentials
 within the WS sphere. However, for a surface calculation the
mean potential at the surface is shifted relative to the bulk and
we need to describe this shift in order to get a good estimate of the
overall potential. At first sight the problem seems hard to overcome
since the mean potential is arbitrary for a bulk calculation due
to the divergence of the electrostatic potential\cite{eldiv}. However
it is possible to circumvent this problem if we use the potential
constructed from a superposition of atom-based densities. We will name
this potential the Harris potential since this is the effective
potential used in the Harris functional, when the input density is obtained
from a
superposition of atom-based densities. We know from previous studies
that when this potential is used as an input to the Harris functional,
excellent
total energies are obtained for phonons, surfaces, and different crystal
structures\cite{chetty,al}.

The electrostatic part of the Harris
potential $V_{el}^{Harris}$ can in a natural way
be divided into a sum over atom-based electrostatic potentials $v_{el}(r)$
each given by the sum of one ionic potential $v_{ion}$ and the Hartree
potential
derived from one atom-based density $\Delta n_{atom}$
\begin{equation}
V_{el}({\bf r}) = \sum_i v_{el}(|{\bf r}-{\bf R}_i|) =
\sum_i (v_{ion}(|{\bf r}-{\bf R}_i|) +
\int \frac{\Delta n_{atom}(|{\bf r}'-{\bf R}_i|)}{|{\bf r}-{\bf r}'|}d{\bf
r}').
\end{equation}
With this construction the 1/r divergence of
the electrostatic potential is avoided,
since the atom-based density decreases exponentially, and thereby fixes the
vacuum level. Note that a consequence of this is that within this approximation
all surfaces of a solid will have the same work function. In contrast to
the total energy the
work function is not variational in the density, and we cannot
expect the density ansatz Eq.~(3) to produce an accurate estimate.

Having established a common energy zero for all Harris potentials we can now
proceed to determine the energy shifts of the mean potential for atoms in
different environments.
In Fig.~3 the solid curve indicates the mean Harris potential for silicon in
the reference
diamond structure as a function of the NS radius. Also shown are the
actual shifts of both the Harris and the self-consistent potential
 at the three principal surfaces and the
potential shifts of the two inequivalent atoms in the LAO(X) phonon.
These potential shifts are marked in the figure at the calculated NS
radii of the atoms. We see that the potential shift in the
reference system compares surprisingly well to the  potential shifts
in the test systems. On the average, the shift of the Harris potential is about
4
percent higher than for the self-consistent potential, and the shift of the
reference system is about 8 percent higher than for the Harris potential.

 In Table~3 we show the r.m.s. error
between the Harris potential and the self-consistent potential for a
silicon atom at the diamond (100) surface. This is compared to the r.m.s
error between the self-consistent surface potential and both the
self-consistent
and Harris potential of the reference
system when the reference system is chosen according to the NS criterion.
As seen from the table the r.m.s errors are similar for
the three potentials, and since the total energies obtained using the
Harris potential are excellent\cite{al,si}, all three
reference potentials are accurate
enough to be used to calculate total energies.

In summary we have presented a scheme for obtaining transferable ASA
potentials. The scheme was applied to six test systems consisting of
silicon or aluminum atoms. The obtained potentials were in very good
agreement with the actual self-consistent potentials, reproducing both
the radial variations  and the shifts in the mean potential,
and when used  as input to the LMTO  method accurate potential
parameters were obtained. We expect the method to be valuable
for constructing new approximate total energy schemes, and
it is presently used  in a new formulation of an Effective-Medium Tight-Binding
model for Silicon\cite{si}.

We are grateful to H. Skriver whose LMTO programs we have used for calculating
the LMTO potential parameters.
We would also like to thank K. Kunc, O.H. Nielsen, R.J. Needs and R.M. Martin
whose solid state programs we have used, and to E.L. Shirley who
developed the pseudo-potential routines. This work was in part supported
by the Danish Research Councils through the Center for Surface
Reactivity. The Center for Atomic-scale Materials Physics is sponsored by
the Danish National Research Foundation. Nithaya Chetty acknowledges
support from the Division of Materials Sciences, U.S.
Department of Energy under contract No. DE--AC02--76CH00016.

\newpage

\begin{figure}
\caption{The figure shows the self-consistent effective potential (i.e.
$r^2v(r)$)
of silicon
in the fcc structure (solid line)  and that
of the diamond reference system with the same NS radius (dotted line)
 and WS radius (dashed line), as a function of the radial distance
The potentials have been aligned such that the mean potential
within the WS sphere is zero. The two vertical lines show the NS and
WS radius of the fcc system, respectively.}
\end{figure}

\begin{figure}
\caption{The figure shows the r.m.s error (Eq.~(2)) between the
effective potential of  silicon in the fcc structure and the potential in the
diamond reference system as a function of the WS radius of the
reference system.The three crosses show the error when the reference
system has the same WS radius, NS radius and bcc WS radius (diamond
with empty spheres) as the fcc system, respectively.}
\end{figure}

\begin{figure}
\caption{The solid curve shows the mean Harris
potential, within the NS, of the  diamond reference
system as a function of the NS radius.
 The two first crosses show the shift in the
mean Harris potential for the two inequivalent atoms in the LAO(X) phonon, at
their NS radii. The last three crosses show the
shift in the mean potential at the diamond (111),
(110), and (100) surface, respectively.
The circles show the shifts for the corresponding self-consistent
potentials. The dotted line shows the mean potential in the equilibrium
diamond lattice.}
\end{figure}

\mediumtext

\begin{table}
\caption{Calculated LMTO potential parameters for the three
pseudopotentials of Fig.~1. For each angular component the
appropriate nonlocal contribution to the potential has been added.}
\begin{tabular}{l|ccc}
potential & $C_p-C_s$ [eV] & $\Delta_s$ [eV] & $\Delta_p$ [eV] \\
\tableline
 SC fcc & 10.968 &  1.566 &  1.358 \\
ref. NS & 10.961 &  1.565 &  1.355  \\
ref. WS & 13.028 &  1.786 &  1.432 \\
\end{tabular}
\end{table}

\begin{table}
\caption{The first row shows the minimal r.m.s. error between the potentials of
the six
test systems and the reference system. The second and third row show the r.m.s.
error when the reference system is that with the same NS and WS radius as the
test
system, respectively. The first three columns show  the  errors for the three
silicon test systems; the fcc
structure, the (100) diamond surface, the diamond longitudinal phonon at the
X point(LAO(X)) with
a displacement of 0.02 in units of the lattice constant. The last three
columns show the  errors for the three aluminum test systems;
 The diamond structure, the fcc (100) surface, and the fcc
longitudinal phonon at the X point(L(X)) with displacement 0.02 in units of
the lattice constant. }
\begin{tabular}{l|ccc|ccc}
& \multicolumn{3}{c|}{Silicon} & \multicolumn{3}{c}{Aluminum} \\
 & FCC  & (100)  & LAO(X)  & Diamond  & (100)  & L(X) \\
\tableline
 $\sigma^{min}$ (eV) & 0.15 & 0.08 & 0.06   & 0.11 & 0.05 & 0.028 \\
 $\sigma^{NS}$ (eV) & 0.15 & 0.20 & 0.08	 & 0.24 & 0.10 & 0.034 \\
 $\sigma^{WS}$  (eV) & 1.49 &1.22 & 0.34	 & 0.89 & 0.48 & 0.042 \\
\end{tabular}
\end{table}

\begin{table}
\caption{The r.m.s. error of potential differences (Eq.~(2))
 for a silicon atom at the (100) surface. The differences are between the
self-consistent surface potential $V^{SC}$, the Harris potential $V^{Harris}$
for the surface with the density construction Eq.~(3),
and the analogous potentials in the diamond reference system chosen according
to the
neutral sphere (NS) criterion.}
\begin{tabular}{lccc}
 \multicolumn{1}{c}{ \ \ } &
\multicolumn{1}{c}{$V^{SC}-V^{Harris}$ } &
\multicolumn{1}{c}{$V^{SC}-V^{SC}_{ref} $} &
\multicolumn{1}{c}{$V^{SC}-V^{Harris}_{ref} $ } \\
\tableline
$\sigma^{NS}$ (eV) & 0.24 & 0.20 & 0.30 \\
\end{tabular}
\end{table}

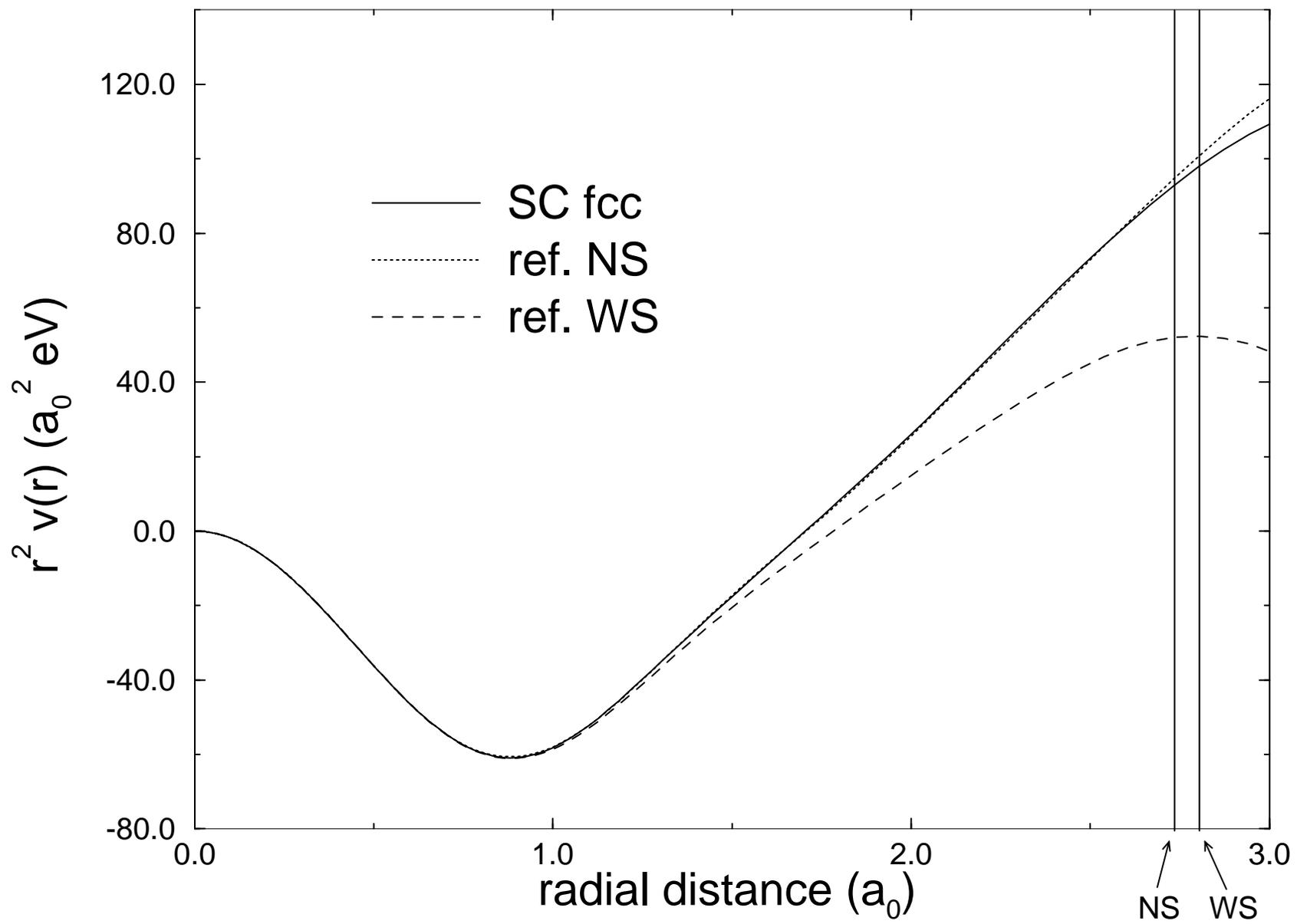
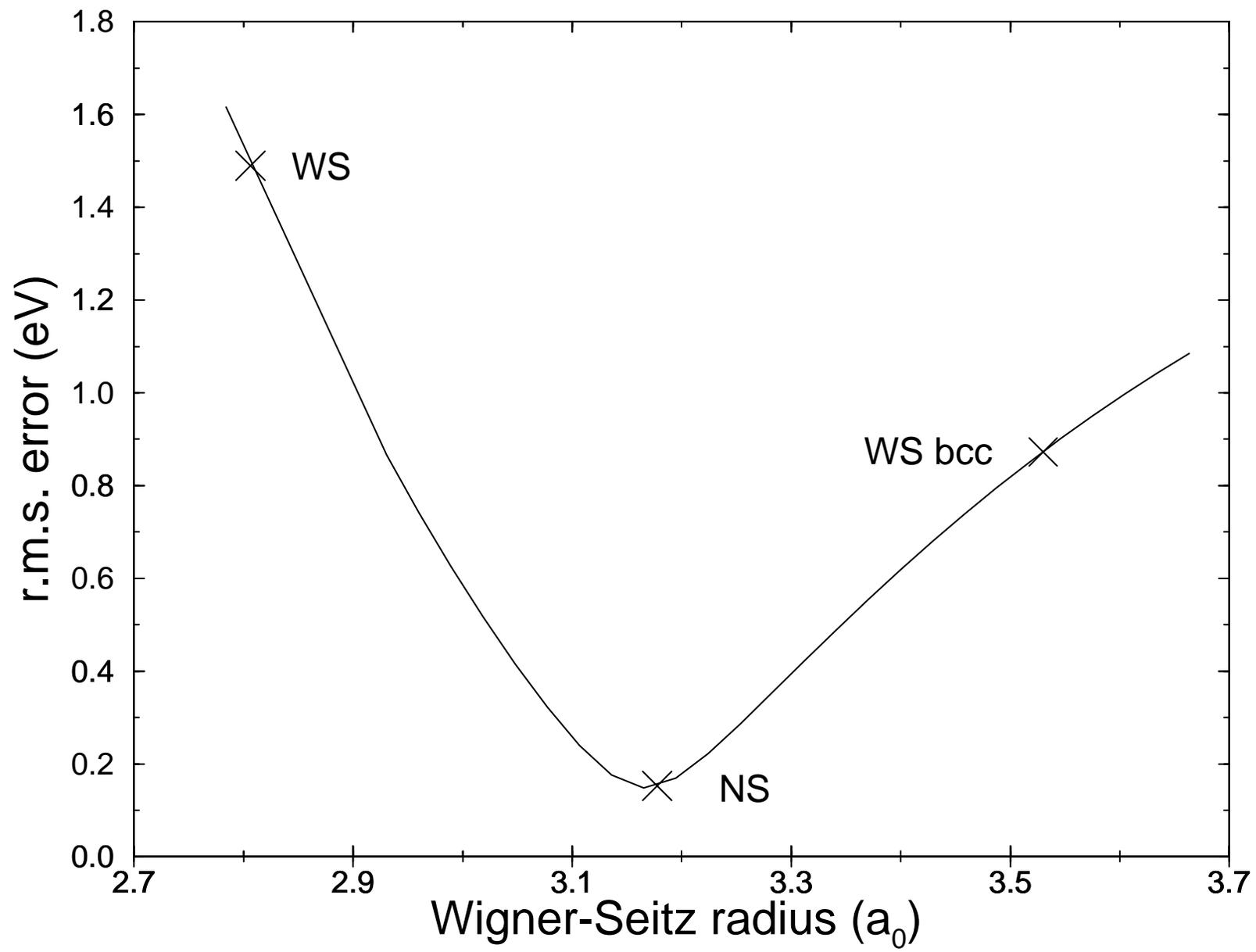
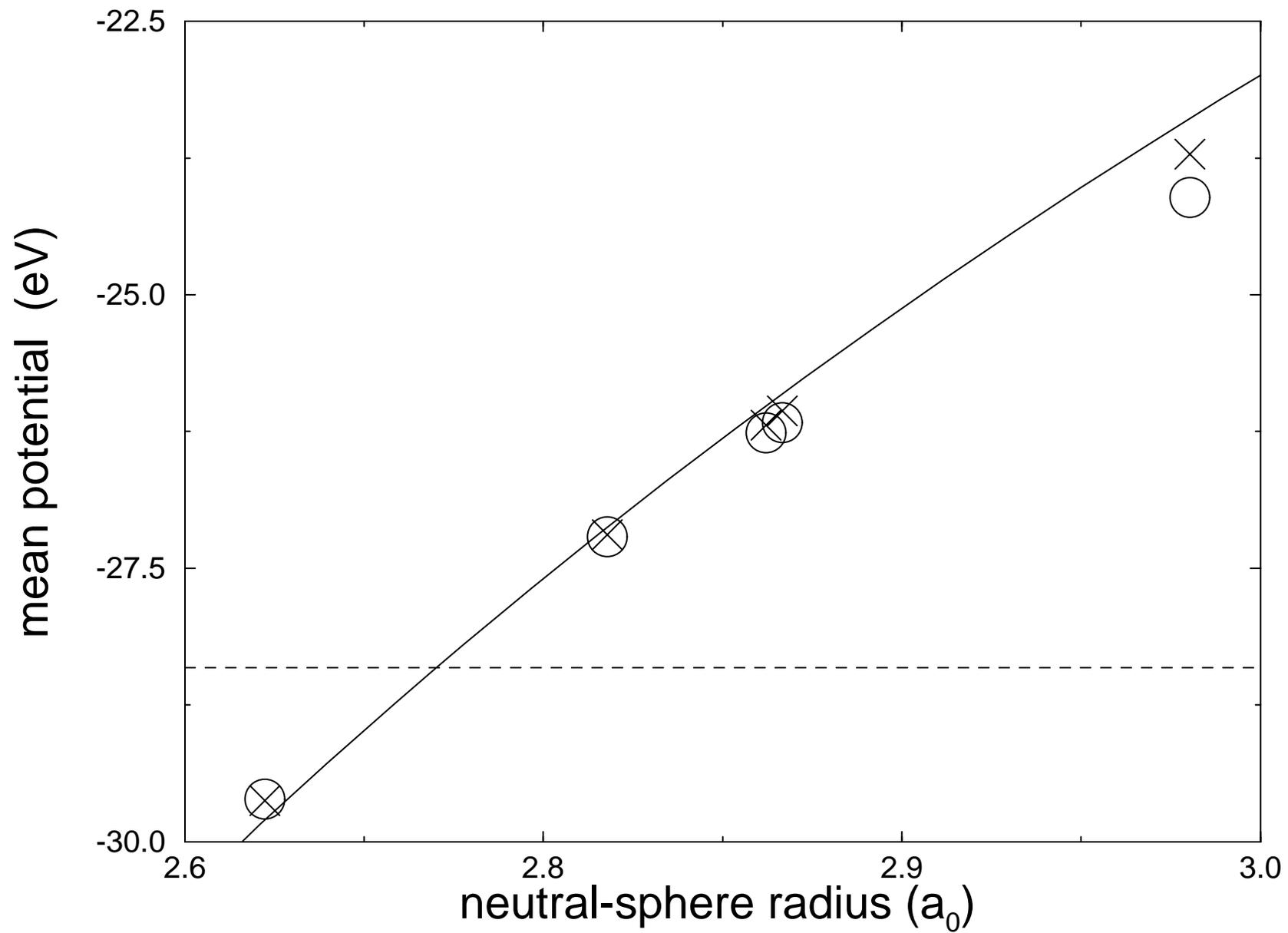
\end{document}